\documentclass[preprint,12pt]{elsarticle}
\usepackage{algorithm}
\usepackage{algorithmicx}
\usepackage{amssymb}
\usepackage{amsmath}

\newcommand{\nnum}{\nonumber}
\newcommand{\EQ}{\begin{eqnarray}}
\newcommand{\EN}{\end{eqnarray}}
\newcommand{\AR}{\begin{array}}
\newcommand{\AN}{\end{array}}

\newtheorem{theorem}{Theorem}
\newtheorem{lemma}{Lemma}
\newtheorem{proof}{Proof}

\begin{document}
\begin{frontmatter}
	
\title{An integrated design of robust decentralized observer and controller for load frequency control\tnoteref{label1}}
\tnotetext[label1]{Xianxian Zhao was supported by the Sustainable Energy Authority of Ireland (22$\backslash$RDD$\backslash$776) and Jianglin Lan was supported by the Leverhulme Trust (ECF-2021-517).}

\author[1]{Xianxian Zhao}
\author[2]{Jianglin Lan\corref{cor1}} 

%% Author affiliation
\affiliation[1]{organization={School of Electrical and Electronic Engineering, University College Dublin},
%	addressline={}, 
	city={Dublin},
	postcode={D04 V1W8}, 
%	state={},
	country={Ireland}}

\affiliation[2]{organization={James Watt School of Engineering, University of Glasgow},
%	addressline={}, 
	city={Glasgow},
	postcode={G12 8QQ}, 
%	state={},
	country={United Kingdom}}

\cortext[cor1]{Corresponding author} 
\ead{Jianglin.Lan@glasgow.ac.uk}

%%%%%%%%%%%%%%%%%%%%%%%%%%%%%%%%%%%%%%%

\begin{abstract}
This paper focuses on designing completely decentralized load frequency control (LFC) for multi-area power systems to achieve global optimized performance. 
To this end, a new concept of integrated design is introduced for designing the decentralized LFC observers and controllers simultaneously off-line, by taking into account of the interactions between areas and the bidirectional effects between the local observer and controller in each area. 
The integrated design in this paper is realized via $H_\infty$ optimization with a single-step linear matrix inequality (LMI) formulation. The LMI regional eigenvalue assignment technique is further incorporated with $H_\infty$ optimization to improve the closed-loop system transient performance.
A three-area power system is simulated to validate the superiority of the proposed integrated design over the conventional decentralized designs.
\end{abstract}

%%Graphical abstract
%\begin{graphicalabstract}
	%\includegraphics{grabs}
%\end{graphicalabstract}

%%Research highlights
%\begin{highlights}
%	\item Research highlight 1
%	\item Research highlight 2
%\end{highlights}

%% Keywords
\begin{keyword}
Load frequency control \sep decentralized control \sep decentralized estimation \sep robust control \sep eigenvalue assignment
\end{keyword}

\end{frontmatter}

\section{Introduction}
 Load frequency control (LFC) is a mechanism for maintaining the power balance between the load demand and the power generation in multi-area interconnected power system. It plays a vital role in keeping the system frequency at its nominal value and maintaining the net tie-line power interchanges between sub-areas at their scheduled values \cite{kundur1994power}. 

There are two LFC stategies: centralized and decentralized. The centralized LFC strategy only has a global controller to control the whole power system, while in the decentralized strategy each sub-area has its own local controller that only uses the state variables within this sub-area. The decentralized strategy is considered to be more economically practical and reliable than the centralized one, especially for large-scale power systems because it can reduce the computational burden and communication complexity \cite{kumar2005recent, shayeghi2009load, pandey2013literature, shankar2017comprehensive}. Thus, this paper focuses on the decentralized LFC strategy. It should be noted that in some works \cite{trinh2013quasi,pham2016load}, the proposed decentralized controllers use state variables from other sub-areas, which are not completely decentralized and do not belong to the decentralized concept discussed in this paper. 

The traditional way of realizing decentralized LFC strategy is to use of Proportional-Integral (PI) controller \cite{bevrani2004robust}, whose parameters are generally obtained through trial-and-error. Many techniques have been published to tune the PI parameters, such as robust control \cite{khodabakhshian2008new, bevrani2004robust}, internal model control \cite{tan2010unified}, genetic algorithms \cite{rerkpreedapong2003robust}, and hybrid evolutionary fuzzy method \cite{juang2006load}. Although the performance can be improved by theses techniques, the PI controllers are static output feedback control with limited control capability and design freedom. 

To overcome the limitations of PI controllers, many advanced control methods using full local state feedback have been proposed for decentralized LFC. 
Paper \cite{tan2012robust} uses robust control to design the local controllers where the tie-line power interchanges are completely ignored. The drawback of their method is that the system with strong tie-line networks can be unstable. 
In some other decentralized methods, e.g., sliding mode control \cite{mi2013decentralized,Su2017fault,rinaldi2017third}, coefficient diagram method \cite{bernard2014decentralized}, active disturbance rejection control \cite{dong2012robust}, and adaptive control \cite{zribi2005adaptive}, the local controllers are designed separately by treating area-interactions (the interactions from other areas acting at one area) as disturbances. This is a passive way of treating the area-interactions in which the sub-areas are not cooperated together to improve the transient response of the whole system. In order to actively include the area-interactions into the decentralized controllers,  \cite{alrifai2011decentralized} and \cite{kazemi2002decentralized} use decomposition methods to obtain the approximation of the area-interactions, and \cite{rerkpreedapong2002decentralized} assumes the area-interactions as white noise and then uses Kalman filter to estimate them. The methods in  \cite{alrifai2011decentralized}, \cite{kazemi2002decentralized} and \cite{rerkpreedapong2002decentralized} have limited application capabilities due to the use of approximation and assumption.

Paper \cite{feliachi1987optimal} uses the optimal control method to design the decentralized LFC. It shows that the centralized control has optimized performance for the whole system with considerations of the area-interactions, and the decentralized controllers can achieve the similar optimized performance of the centralized one. 
However, the control method proposed in \cite{feliachi1987optimal} has three drawbacks: 1) the decentralized controllers are feasible only under certain conditions; 2) the design is based on the steady state situation; 3) the model of each sub-area does not include the tie-line power flow. 

Aiming to achieve optimized performance for the whole power system, this paper proposes a new design concept that the completely decentralized LFC controllers are designed simultaneously from the view of the whole power system in which the area-interactions are considered.  

Most of the papers for decentralized LFC assume that all the state variables are known \cite{feliachi1987optimal, tan2012robust, mi2013decentralized, bernard2014decentralized, zribi2005adaptive}. In order to be more economically practical, observers, e.g. adaptive observer \cite{kazemi2002decentralized}, third-order sliding mode observer \cite{rinaldi2017third}, PI sliding mode observer \cite{Su2017fault}, and Kalman filter \cite{rerkpreedapong2002decentralized}, have been used to estimate the state variables.  
However, the observers in \cite{kazemi2002decentralized, rinaldi2017third, rerkpreedapong2002decentralized} are designed separately from the controllers by assuming that the estimation errors are zero when proceeding the controller design. 
In this way, the acceptable transient performance of the closed-loop control systems cannot be guaranteed. To overcome this limitation, \cite{Su2017fault} proposes a two-step approach to design the observer first and then the controller by considering the estimation error. However, in \cite{Su2017fault} the effects of the control system on the observer are not taken into account. For an observer-based control system there exist bidirectional effects between the observer and the controller: 1) the estimation errors affect the control performances, and 2) the control performances have effects on the observer, which should be taken into account.

Considering the above background, this paper has the following contributions when compared with the existing decentralized LFC approaches:

$\bullet$ For the first time, an integrated design concept is proposed for designing all the decentralized observers and controllers simultaneously for LFC. This design strategy takes into account of both 1) the area-interactions and 2) the bidirectional effects between the local observers and controllers, and it thus achieves global optimized LFC performance. 

$\bullet$ The proposed robust observer-based full-state feedback controllers are completely decentralized and their gains are solved off-line via $H_\infty$ optimization using an single-step LMI approach. Although the decentralized observers and controllers are all designed together, they are completely decentralized in implementation, i.e., each sub-area has its own local observer and controller.   

$\bullet$ The transient performance of the
observer-based control system is further improved by applying the LMI regional eigenvalue assignment technique.  

The paper is organized as follows. Section \ref{Power system model} describes the state-space model of a $\mathrm{N}$-area power system. Section \ref{decentralized design} presents the integrated design of the decentralized observer and controller. Section \ref{simulation} provides a three-area power system as a demonstrative example and Section \ref{conclusion} draws a conclusion.

Notation: The symbol $\mathbb{R}^n$ represents the $n$ dimension real number space, $\| \cdot \|_\infty$ represents the $\infty$-norm in the Euclidean space, $I$ is an identity matrix of appropriate dimension, $\mathrm{He}(W) = W + W^\top$, $\star$ represents the transpose of the element at its symmetric position in a matrix, and $\left[\Omega_{i,j} \right]_{n \times n}$ represents that $\Omega$ is a $n \times n$ dimension block symmetric matrix whose $(i,j)$ element is defined as $\Omega_{i,j}$.

\section{Multi-area power system model} \label{Power system model}
\begin{figure}[H]
	\centering
	\includegraphics[width=0.8\columnwidth]{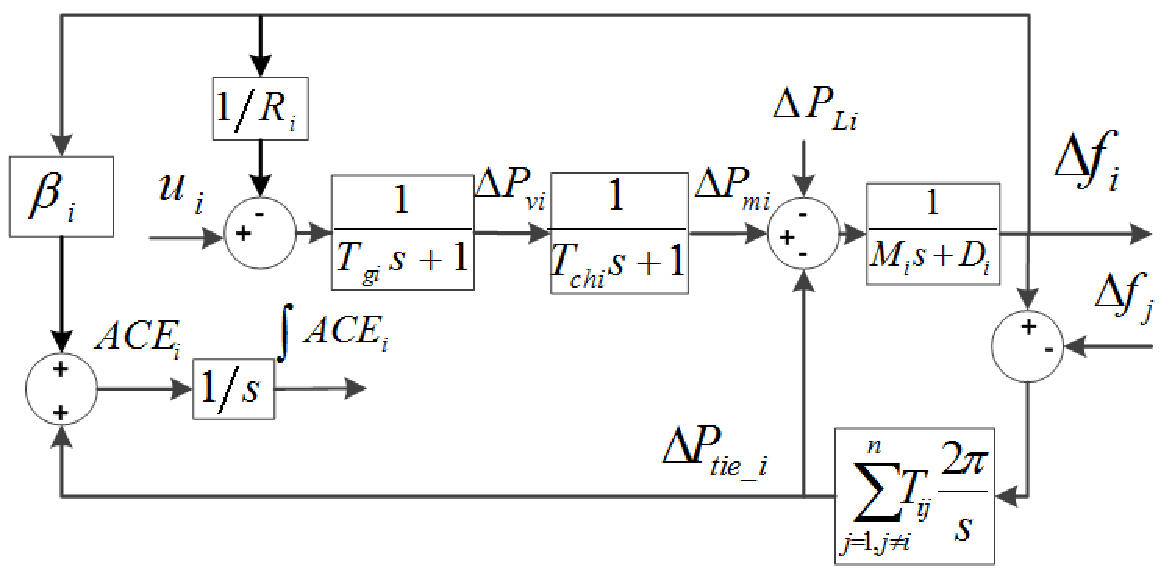}    
	\caption{Dynamic model of the $i$th area in a $\mathrm{N}$-area LFC scheme}
	\label{fig1}
\end{figure}

This paper considers a linearized model similar to the model used in \cite{jiang2012delay}. Fig. \ref{fig1} shows the $i$th area of the $\mathrm{N}$-area power system, whose state-space model is represented by  
\EQ \label{sys i}
\dot{x}_i &=& A_i x_i + \sum_{j=1,j\neq i}^{N} \Delta A_{ij} x_j + B_i u_i + F_i d_i \nnum\\
y_i &=& C_i x_i,
\EN
where 
\EQ
&& x_i = \left[ \Delta f_i ~ \Delta P_{mi} ~ \Delta P_{vi}~  \Delta P_{tiei} ~\int ACE_i \right]^\top, \nnum\\
&& ACE_i = \beta_i \Delta f_i + \Delta P_{tiei}, ~d_i = \Delta P_{Li}, \nnum\\
&& A_i = \left[\AR{ccccc}  
-\frac{D_i}{M_i} & \frac{1}{M_i} & 0 & - \frac{1}{M_i} & 0 \\
0 & - \frac{1}{T_{chi}} & \frac{1}{T_{chi}} & 0 & 0 \\
- \frac{1}{R_i T_{gi}} & 0 & - \frac{1}{T_{gi}} & 0 & 0 \\
2 \pi \sum_{j=1,j\neq i}^{N}T_{ij} & 0 & 0 & 0 & 0\\
\beta_i & 0 & 0 & 1 & 0
\AN\right], ~\nnum\\
&& B_i = \left[\AR{c} 0 \\ 0 \\ \frac{1}{T_{gi}} \\ 0  \\ 0\AN \right], ~
F_i = \left[\AR{c} -\frac{1}{M_i} \\ 0 \\ 0 \\ 0 \\ 0 \AN \right], ~C_i = \left[\AR{ccccc} 1 & 0 & 0 & 0 & 0 \\ 0 & 0 & 0 & 1 & 0 \\ 0 & 0 & 0 & 0 & 1 \AN \right],\nnum \\
&& \Delta A_{ij} = \left[\AR{ccccc}  
0 & 0 & 0 & 0 & 0\\
0 & 0 & 0 & 0 & 0\\
0 & 0 & 0 & 0 & 0\\
-2\pi T_{ij} & 0 & 0 & 0 & 0\\
0 & 0 & 0 & 0 & 0
\AN\right], ~T_{ij}=T_{ji}. \nnum
\EN
$x_i$ is the state vector; $u_i$ is the control input; $d_i$ is the vector of load disturbance; $y_i$ is the output vector; $\Delta f_i$, $\Delta P_{mi}$, $\Delta P_{vi}$ and $\Delta P_{Li}$ are the deviations of frequency, the generator mechanical output, valve position, and load, respectively. $M_i$, $D_i$, $T_{gi}$, $T_{chi}$, and $R_i$ denote the moment of inertia of the generator, generator damping coefficient, time constant of the governor, time constant of the turbine, and speed drop, respectively; $\beta_i$ is frequency bias factor; $T_{ij}$ is the tie-line synchronizing coefficient between the $i$th and $j$th control area; $\sum_{j=1,j\neq i}^{N} \Delta A_{ij} x_j$ are the area-interactions. $A_i \in \mathbb{R}^{n_i \times n_i}$, $\Delta A_{ij} \in \mathbb{R}^{n_i \times n_j}$, $B_i \in \mathbb{R}^{n_i \times m_i}$, $F_i \in \mathbb{R}^{n_i \times q_i}$, and $C_i \in \mathbb{R}^{p_i \times n_i}$ are known constant matrices.

According to (\ref{sys i}), the state-space model of the $\mathrm{N}$-area power system can be represented as
\EQ \label{sys}
\dot{x} &=& A x + \Delta A x + B u + F d \nnum\\
y &=& C x,
\EN
where 
\begin{align}
x &= \left[ x_1^\top~ x_2^\top ~\cdots ~x_N^\top \right]^\top,~ u = \left[ u_1^\top ~u_2^\top ~\cdots~ u_N^\top \right]^\top, \nnum\\
d &= \left[ d_1^\top~ d_2^\top~ \cdots~ d_N^\top \right]^\top,~ y = \left[ y_1^\top~ y_2^\top ~\cdots~ y_N^\top \right]^\top, \nnum\\
A &= \mathrm{diag} (A_1, A_2, \cdots, A_N),~ B = \mathrm{diag} (B_1, B_2, \cdots, B_N), \nnum\\
F &= \mathrm{diag} (F_1, F_2, \cdots, F_N),~ C = \mathrm{diag} (C_1, C_2, \cdots, C_N), \nnum\\
\Delta A &= \left[\AR{cccc} 0 & \Delta A_{12} & \cdots & \Delta A_{1N} \\
\Delta A_{21} & 0 & \cdots & \Delta A_{2N} \\
\vdots& \vdots & \ddots &\vdots \\
\Delta A_{N1} &\Delta A_{N2} &\cdots & 0 \AN\right]. \nnum
\end{align}
Not that $A \in \mathbb{R}^{n \times n}$, $\Delta A \in \mathbb{R}^{n \times n}$, $B \in \mathbb{R}^{n \times m}$, $F \in \mathbb{R}^{n \times q}$, and $C \in \mathbb{R}^{p \times n}$ are known constant matrices, where $n = \sum_{i=1}^{N} n_i$, $m = \sum_{i=1}^{N} m_i$, $q = \sum_{i=1}^{N} q_i $, and $p = \sum_{i=1}^{N} p_i$.

\section{Integrated design of decentralized observer and controller} \label{decentralized design}
This section describes the proposed integrated strategy of designing together all the decentralized observers and controllers based on $H_\infty$ optimization using a single-step LMI formulation. 

\subsection{Observer and controller structure}
The decentralized observer for the $i$th area is designed as  
\EQ \label{observer i}
\dot{z}_i &=& \varPhi_i z_i + G_i u_i + L_i y_i \nnum\\
\hat{x}_i &=& z_i + H_i y_i,
\EN
where $z_i \in \mathbb{R}^{n_i}$ is the observer system state vector and $\hat{x}_i \in \mathbb{R}^{n_i}$ is the estimate of $x_i$. $\varPhi_i \in \mathbb{R}^{n_i \times n_i}$, $G_i \in \mathbb{R}^{n_i \times m_i}$, $L_i \in \mathbb{R}^{n_i \times p_i}$, and $H_i \in \mathbb{R}^{n_i \times p_i}$ are the observer design gains. 

According to (\ref{observer i}), the composite form of all the $N$ decentralized observers is represented by
\EQ \label{observer}
\dot{z} &=& \varPhi z + G u + L y \nnum\\
\hat{x} &=& z + H y,
\EN
where $z = [z_1^\top ~ z_2^\top ~ \cdots ~ z_N^\top]^\top$ and $\hat{x} = [\hat{x}_1^\top ~ \hat{x}_2^\top ~ \cdots ~ \hat{x}_N^\top]^\top$. The matrices $\varPhi = \mathrm{diag} (\varPhi_1, \varPhi_2, \cdots, \varPhi_N)$, $G = \mathrm{diag} (G_1, G_2, \cdots, G_N)$, $L = \mathrm{diag} (L_1, L_2, \cdots, L_N)$, and $H = \mathrm{diag} (H_1, H_2, \cdots, H_N)$ are the observer gains to be designed. 

Define the estimation error as $e = x - \hat{x}$, it follows from (\ref{sys}) and (\ref{observer}) that
\EQ \label{error dynamic0}
\dot{e} &=& \Psi \dot{x} - \dot{z} \nnum\\
&=& (\Psi A - L_1 C) e + (\Psi A - L_1 C - \varPhi) z + (\Psi B  - G) u  \nnum\\
&& + [(\Psi A - L_1 C)H - L_2 ] y + \Psi \Delta A x + \Psi F d,
\EN
where $\Psi = I_n - H C$ and $L = L_1 + L_2$. 

Define the following matrix equations:
\EQ
\label{cond1}
\Psi A - L_1 C - \varPhi &=& 0, \\
\label{cond2}
\Psi B - G &=& 0, \\
\label{cond3}
(\Psi A - L_1 C) H - L_2 &=& 0, \\
\label{cond4}
\Psi F &=& 0.
\EN

Substituting (\ref{cond1}) - (\ref{cond3}) into (\ref{error dynamic0}) gives
\EQ \label{error dynamic1}
\dot{e} = (\Psi A - L_1 C) e + \Psi \Delta A x.
\EN

It can be seen from (\ref{cond1}) - (\ref{cond4}) that once the matrices $L_1$ and $H$ are obtained, then all the other matrices $\varPhi$, $G$, and $L_2$ can be calculated from (\ref{cond1}) - (\ref{cond4}). Since $\mathrm{rank}(C F) = \mathrm{rank}(F) = q$, there always exists a matrix $H$ such that (\ref{cond4}) is satisfied, which leads to the total decoupling of the load disturbance $d$ \cite{patton1997observer}. The solution of $H$ is $H = F [(CF)^\top (CF)]^{-1} (CF)^\top$. The matrix $L_1$ is designed in Section \ref{gains determination} such that the error system (\ref{error dynamic1}) is robustly asymptotically stable.

The decentralized full-state feedback controller for the $i$th area is designed as
\EQ \label{controller i}
u_i = - K_i \hat{x}_i,
\EN
where $K_i \in \mathbb{R}^{m_i \times n_i}$ is the controller gain to be designed.

According to (\ref{controller i}), the composite form of all the $N$ decentralized controllers is represented by 
\EQ \label{controller}
u = - K \hat{x},
\EN
where $K = \mathrm{diag}(K_1, K_2, \cdots, K_N)$ is the controller gain to be determined.

Substituting the controller (\ref{controller}) into the system (\ref{sys}) gives the closed-loop system
\EQ \label{closed-sys1}
\dot{x} = (A - B K) x + B K e + \Delta A x + F d.
\EN

The composite closed-loop system consisting of (\ref{error dynamic1}) and (\ref{closed-sys1}) is
\EQ \label{closed-sys2}
\dot{x} &=& (A - B K) x + B K e + \Delta A x + F d \nnum\\
\dot{e} &=& (\Psi A - L_1 C) e + \Psi \Delta A x \nnum\\
z_c &=& C_x x + C_e e,
\EN
where $z_c \in \mathbb{R}^n$ is the measured output to verify the control and estimation performances. $C_x \in \mathbb{R}^n$ and $C_e \in \mathbb{R}^n$ are given constant matrices. In (\ref{closed-sys2}) only the gains $K$ and $L_1$ need to be determined, which is provided in Section \ref{gains determination}.

It should be noted that different from the separated decentralized LFC strategy used in the literature, e.g., \cite{kazemi2002decentralized, rinaldi2017third, rerkpreedapong2002decentralized, Su2017fault}, in the proposed strategy the area-interactions $\Delta A x$ are treated as system uncertainty, and the bidirectional effects ($B K e$ and $\Psi \Delta A x$) between the observer and the control system are taken into account.

\subsection{Robust performance analysis and gains determination} \label{gains determination}
This section analyzes the robust performance of the composite closed-loop system (\ref{closed-sys2}) using Theorem \ref{theorem1}, and provides the LMI for solving the gains $K$ and $L_1$ in Theorem \ref{theorem2}.

\begin{theorem} \label{theorem1}
	Given positive scalars $\gamma$ and $\varepsilon_1$, the composite closed-loop system (\ref{closed-sys2}) is stable with $H_\infty$ performance $\| G_{z_c d} \|_{\infty} < \gamma$, if there exist symmetric positive definite matrices $P \in \mathbb{R}^{n \times n}$ and $Q \in \mathbb{R}^{n \times n}$, and matrices $H \in \mathbb{R}^{n \times p}$ and $L_1 \in \mathbb{R}^{n \times p}$ such that
	\EQ \label{theorem1 eq1}
	\left[\AR{cccc}  
	\Pi_{1,1} & P B K & P F & C_x^\top \\
	\star & \Pi_{2,2} & 0 & C_e^\top \\
	\star & \star & - \gamma^2 I & 0 \\
	\star & \star & \star & -I
	\AN\right] < 0, 
	\EN
	where $\Pi_{1,1} = \mathrm{He} \left[ P (A - B K) + P \Delta A \right] + \varepsilon_1^{-1} \Delta A^\top \Delta A$ and $\Pi_{2,2} = \mathrm{He} \left[ Q (\Psi A - L_1 C) \right] + \varepsilon_1 Q \Psi \Psi^\top Q$.
\end{theorem}
\begin{proof}	
	Consider a Lyapunov functioin $V_x = x^\top P x$. Its time derivative is
	\EQ \label{dot Vx}
	\dot{V}_x = x^\top \mathrm{He} \left[ P (A - B K) + P \Delta A \right] x + \mathrm{He} (x^\top P B K e)  + \mathrm{He} (x^\top P F d).
	\EN
	
	Consider another Lyapunov functioin $V_e = e^\top Q e$. Its time derivative is
	\EQ \label{dot Ve}
	\dot{V}_e = e^\top \mathrm{He} \left[ Q (\Psi A - L_1 C) \right] e + \mathrm{He} (e^\top Q \Psi \Delta A x).
	\EN
	
	There exists a constant $\varepsilon_1$ satisfying that
	\EQ \label{Uncertainty}
	\mathrm{He} (e^\top Q \Psi \Delta A x) \leq \varepsilon_1 e^\top Q \Psi \Psi^\top Q e + \varepsilon_1^{-1} x^\top \Delta A^\top \Delta A x. 
	\EN
	
	Substituting (\ref{Uncertainty}) into (\ref{dot Ve}) yields
	\EQ \label{dot Ve 2}
	\dot{V}_e \leq e^\top \left\{ \mathrm{He} \left[ Q (\Psi A - L_1 C) \right] + \varepsilon_1 Q \Psi \Psi^\top Q \right\} e  + \varepsilon_1^{-1} x^\top \Delta A^\top \Delta A x.
	\EN
	
	The $H_\infty$ performance $\| G_{z_c d} \|_\infty < \gamma$ can be represented as
	\EQ \label{J1}
	J = \int_0^{\infty} \left( z_c^\top z_c - \gamma^2 d^\top d \right) dt < 0. 
	\EN
	Under zero initial conditions (i.e., $V_x(0) = 0$ and $V_e(0) = 0$),
	\EQ
	J &=& \int_0^{\infty} ( z_c^\top z_c - \gamma^2 d^\top d + \dot{V}_x + \dot{V}_e ) dt - \int_0^{\infty} (\dot{V}_x + \dot{V}_e) dt \nnum\\
	&=& \int_0^{\infty} ( z_c^\top z_c - \gamma^2 d^\top d + \dot{V}_x + \dot{V}_e ) dt 
	 - ( V_x(\infty) + V_e(\infty) ) + ( V_x(0) + V_e(0) )\nnum\\
	&\leq& \int_0^{\infty} ( z_c^\top z_c - \gamma^2 d^\top d + \dot{V}_x + \dot{V}_e ) dt. \nnum
	\EN
	
	Hence, a sufficient condition for (\ref{J1}) is
	\EQ \label{J}
	z_c^\top z_c - \gamma^2 d^\top d + \dot{V}_x + \dot{V}_e < 0.
	\EN
	
	Define $\eta = [x^\top ~ e^\top ~ d^\top]^\top$. Substituting (\ref{dot Vx}) and (\ref{dot Ve 2}) into (\ref{J}) and using the Schur Complement \cite{boyd1994linear} gives (\ref{theorem1 eq1}). \qed
\end{proof}

Since the inequality (\ref{theorem1 eq1}) in Theorem \ref{theorem1} is bilinear matrix inequality and cannot be solved directly via Matlab LMI toolbox, it is further converted into a LMI condition (\ref{theorem2 eq1}) presented in Theorem \ref{theorem2}, which is used to solve the gains $K$ and $L_1$.

\begin{theorem} \label{theorem2}
	Given positive scalars $\gamma$, $\varepsilon_1$, and $\varepsilon_2$, the composite closed-loop system (\ref{closed-sys2}) is stable with $H_\infty$ performance $\| G_{z_c d} \|_{\infty} < \gamma$, if there exist symmetric positive definite matrices $Z_i \in \mathbb{R}^{n_i \times n_i}$ and $Q_i \in \mathbb{R}^{n_i \times n_i}$, and matrices $M_{1i} \in \mathbb{R}^{m_i \times n_i}$ and $M_{2i} \in \mathbb{R}^{n_i \times p_i}$, $i = 1, 2, \dots, N$,  such that
	\EQ \label{theorem2 eq1}
	\left[\Omega_{i,j} \right]_{8 \times 8} < 0,
	\EN
	where $\Omega$ is a $8 \times 8$ symmetric block matrix whose elements $\Omega_{i,j}$ are defined as follows:
	For $1 \leq i \leq j \leq 8$, 
	\EQ
	&&\Omega_{1,1} = \mathrm{He}(\Omega_{1,1}^1 + \Omega_{1,1}^2), \nnum\\
	&&\Omega_{1,1}^1 = \mathrm{diag}(A_1 Z_1 - B_1 M_{11}, A_2 Z_2 - B_2 M_{12}, \cdots, A_N Z_N - B_N M_{1N}), \nnum \\ 
	&&\Omega_{1,1}^2 = \left[\AR{cccc} 0 & \Delta A_{12} Z_2 & \cdots & \Delta A_{1N} Z_N \\
	\Delta A_{21} Z_1 & 0 & \cdots & \Delta A_{2N} Z_N \\
	\vdots& \vdots & \ddots &\vdots \\
	\Delta A_{N1} Z_1 &\Delta A_{N2} Z_2 &\cdots & 0 \AN\right],\nnum\\
	&&\Omega_{1,3} = \mathrm{diag}(F_1, F_2, \cdots, F_N),~
	\Omega_{1,4} = \mathrm{diag}(Z_1 C_{x_1}^\top, Z_2 C_{x_2}^\top, \cdots, Z_N C_{x_N}^\top), \nnum\\
	&&\Omega_{1,5} = (\Omega_{1,1}^2)^\top,~
	\Omega_{1,7} = \mathrm{diag}(B_1 M_{11}, B_2 M_{12}, \cdots, B_N M_{1N}),\nnum \\
	&&\Omega_{2,2} = \mathrm{He}(\mathrm{diag}(Q_1 \Psi_1 A_1 - M_{21} C_1, Q_1 \Psi_2 A_2 - M_{22} C_2,\nnum \\
	&&\hspace*{1.1cm} \cdots, Q_N \Psi_N A_N - M_{2N} C_N)),\nnum \\
	&&\Omega_{2,4} = \mathrm{diag}(C_{e_1}^\top, C_{e_2}^\top, \cdots, C_{e_N}^\top),\nnum \\
	&&\Omega_{2,6} = \mathrm{diag}(Q_1 \Psi_1, Q_2 \Psi_2, \cdots, Q_N \Psi_N), \Omega_{2,8} = I, \nnum\\
	&&\Omega_{3,3} = - \gamma^2 I, \Omega_{4,4} = - I, \Omega_{5,5} = - \varepsilon_1 I, \Omega_{6,6} = - \varepsilon_1^{-1} I,\nnum\\
	&&\Omega_{7,7} = - \varepsilon_2^{-1} Z, \Omega_{8,8} = - \varepsilon_2 Z,\nnum 
	\EN
	and all the other elements are zero.
	Then the observer and controller gains are given by: $K_i = M_{1i} Z_i^{-1}$, $L_{i1} = Q_i^{-1} M_{2i}$, $i = 1, 2, \dots, N$.
\end{theorem}

\begin{proof}
	Define $Z = P^{-1}$. Pre- and post-multiplying both sides of (\ref{theorem1 eq1}) by $\mathrm{diag}(Z, I, I, I)$ and its transpose gives
	\EQ \label{theorem1 eq1 2}
	\left[\AR{cccc}  
	\Pi_{1,1}^1 & B K & F & Z C_x^\top \\
	\star & \Pi_{2,2}^1 & 0 & C_e^\top \\
	\star & \star & - \gamma^2 I & 0 \\
	\star & \star & \star & -I
	\AN\right] < 0, 
	\EN
	where 
	\begin{align}
		\Pi_{1,1}^1 &= \mathrm{He} \left[ (A - B K) Z + \Delta A Z \right] + \varepsilon_1^{-1} Z \Delta A^\top \Delta A Z, \nnum \\
		\Pi_{2,2}^1 &= \mathrm{He} \left[ Q (\Psi A - L_1 C) \right] + \varepsilon_1 Q \Psi \Psi^\top Q. \nnum
	\end{align}
	
	By applying Schur Complement \cite{boyd1994linear} to (\ref{theorem1 eq1 2}), we can linerize the terms $\varepsilon_1^{-1} Z \Delta A^\top \Delta A Z$ and $\varepsilon_1 Q \Psi \Psi^\top Q$, which gives
	\EQ \label{theorem1 eq1 3}
	\left[\AR{cccccc}  
	\Pi_{1,1}^2 & B K & F & Z C_x^\top & Z \Delta A^\top & 0\\
	\star & \Pi_{2,2}^2 & 0 & C_e^\top & 0 & Q \Psi \\
	\star & \star & - \gamma^2 I & 0 & 0 & 0 \\
	\star & \star & \star & -I & 0 & 0 \\
	\star & \star & \star & \star & -\varepsilon_1 I & 0 \\
	\star & \star & \star & \star & \star & - \varepsilon_1^{-1} I
	\AN\right] < 0, 
	\EN
	where $\Pi_{1,1}^2 = \mathrm{He} \left[ (A - B K) Z + \Delta A Z \right]$ and $\Pi_{2,2}^2 = \mathrm{He} \left[ Q (\Psi A - L_1 C) \right]$.
	
	According the Young Inequality \cite{boyd1994linear}, there exists a constant $\varepsilon_2$ such that
	\EQ \label{BK}
	\mathrm{He} (\Gamma_1^\top \Gamma_2) \leq \varepsilon_2 (\Gamma_1^\top Z) Z^{-1} (Z \Gamma_1) + \varepsilon_2^{-1} \Gamma_2^\top Z^{-1}\Gamma_2,
	\EN
	where $\Gamma_1 = \left[(B K)^\top ~ 0 ~ 0 ~ 0 ~ 0 ~ 0 \right]^\top$ and $\Gamma_2 = \left[0 ~ I ~ 0 ~ 0 ~ 0 ~ 0\right]$.
	
	Substituting (\ref{BK}) into (\ref{theorem1 eq1 3}) gives
	\EQ \label{theorem1 eq1 4}
	\left[\Pi_{i,j}^3 \right]_{8 \times 8} < 0, 
	\EN
	where $\Pi^3$ is $8 \times 8$ symmetric block matrix whose elements $\Pi_{i,j}^3$ are defined as follows:
	For $1 \leq i \leq j \leq 8$, 
	$\Pi_{1,1}^3 =\mathrm{He} \left[ (A - B K) Z + \Delta A Z \right], \Pi_{1,3}^3 = F, \Pi_{1,4}^3 = Z C_x^\top,$ 
	$\Pi_{1,5}^3 = Z \Delta A^\top, \Pi_{1,7}^3 = BKZ, \Pi_{2,2}^3 = \mathrm{He} \left[ Q (\Psi A - L_1 C) \right],$ 
	$\Pi_{2,4}^3 = C_e^\top, \Pi_{2,6}^3 = Q \Psi, \Pi_{2,8}^3 = I, \Pi_{3,3}^3 = - \gamma^2 I, \Pi_{4,4}^3 = -I,$ 
	$\Pi_{5,5}^3 = -\varepsilon_1 I, \Pi_{6,6}^3 = -\varepsilon_1^{-1} I, \Pi_{7,7}^3 = - \varepsilon_2^{-1} Z, \Pi_{8,8}^3 = - \varepsilon_2 Z,$ 
	and all the other elements are zero.
	
	Define the following matrices:
	\begin{align}
		Z &= \mathrm{diag}(Z_1, Z_2, \cdots, Z_N), ~Q = \mathrm{diag}(Q_1, Q_2, \cdots, Q_N), \nnum\\
		K &= \mathrm{diag}(K_1, K_2, \cdots, K_N),~H = \mathrm{diag}(H_1, H_2, \cdots, H_N), \nnum\\
		\Psi &= \mathrm{diag}(\Psi_1, \Psi_2, \cdots, \Psi_N),~L_1 = \mathrm{diag}(L_{11}, L_{21}, \cdots, L_{N1}), \nnum\\
		C_x &= \mathrm{diag}(C_{x1}, C_{x2}, \cdots, C_{xN}),~ C_e = \mathrm{diag}(C_{e1}, C_{e2}, \cdots, C_{eN}), \nnum\\
		M_{1i} &= K_i Z_i, ~M_{2i} = Q_i L_{i1}, ~i = 1, 2, \dots, N. \nnum
	\end{align}
	Substituting these matrices into (\ref{theorem1 eq1 4}) gives (\ref{theorem2 eq1}). \qed
\end{proof}

\subsection{Time response improvement by eigenvalue assignment} \label{time response}
Although Theorems \ref{theorem1} and \ref{theorem2} based on $H_\infty$ optimization can ensure that the composite closed-loop system (\ref{closed-sys2}) is robustly asymptotically stable, they have no guarantee of suitable time response of the closed-loop system. To tackle with the above issue, this section uses the LMI regional eigenvalue assignment technique in Lemma \ref{lemma pole} together with Theorem \ref{theorem2} to guarantee acceptable closed-loop system time response. By using the techniques described in \cite{chilali1996h}, the eigenvalues can be assigned to different regions, e.g., vertical strips, disks and conic sectors. This paper, however, considers the vertical strips as a tutorial example to show that the eigenvalue assignment technique can be combined with $H_\infty$ optimization to improve the transient performance of the closed-loop LFC system. 
\begin{lemma}\cite{chilali1996h} \label{lemma pole}
	The system $\dot{x} = A x$ is assigned to is a vertical strip region: $a < \mathrm{Re}(\lambda) < b$ with $\lambda$ denotes the eigenvalues of $A$, if there exists a symmetric positive definite matrix $P_0$ such that
	\EQ
	\left[\AR{cc} \mathrm{He}(P_0 A) - 2 b P_0 & 0 \\
	0 & - \mathrm{He}(P_0 A) + 2 a P_0 \AN\right] < 0. \nnum
	\EN
\end{lemma}

Following Theorem \ref{theorem2} and Lemma \ref{lemma pole}, Theorem \ref{theorem3} is proposed to ensure the composite closed-loop system (\ref{closed-sys2}) is 1) robustly asymptotically stable, and 2) with eigenvalues assigned into prescribed strip regions.

The strip regions are designed as $\mathcal{D}_{j,i} = \{\lambda_{j,i}: a_{j,i} < \mathrm{Re}(\lambda_{j,i}) < b_{j,i}\}$, $j = 1, 2$, $i = 1, 2, \dots, N$. $\lambda_{1,i}$ and $\lambda_{2,i}$ are the eigenvalues of the control system matrix and the observer system matrix, respectively. $a_{j,i}$ and $b_{j,i}$ are negative design constants.  

\begin{theorem} \label{theorem3}
	Given positive scalars $\gamma$, $\varepsilon_1$, and $\varepsilon_2$, and the constant matrices  
	$a_1 = \mathrm{diag}(a_{1,1}, \cdots, a_{1,N})$, 
	$b_1 = \mathrm{diag}(b_{1,1}, \cdots, b_{1,N})$, 
	$a_2 = \mathrm{diag}(a_{2,1}, \cdots, a_{2,N})$, and
	$b_2 = \mathrm{diag}(b_{2,1}, \cdots, b_{2,N})$. The composite closed-loop system (\ref{closed-sys2}) is 1) stable with $H_\infty$ performance $\| G_{z_c d} \|_{\infty} < \gamma$ and 2) the eigenvalues of its system matrix are assigned to the regions $\mathcal{D}_{j,i}$, if there exist symmetric positive definite matrices $Z_i \in \mathbb{R}^{n_i \times n_i}$ and $Q_i \in \mathbb{R}^{n_i \times n_i}$, and matrices $M_{1i} \in \mathbb{R}^{m_i \times n_i}$ and $M_{2i} \in \mathbb{R}^{n_i \times p_i}$, $i = 1, 2, \dots, N$,  such that
	\EQ \label{theorem3 eq1}
	\left[\Omega_{i,j} \right]_{8 \times 8} &< 0, \\
	\label{theorem3 eq2}
\left[\AR{cc} \Omega_{1,1} - 2 b_1 Z & 0 \\
	\star & -\Omega_{1,1} + 2 a_1 Z
	\AN\right] &< 0, \\
	\label{theorem3 eq3}
	\left[\AR{cc} \Omega_{2,2} - 2 b_2 Q & 0 \\ 
	\star & - \Omega_{2,2} + 2 a_2 Q  
	\AN\right] &< 0,
	\EN
	where $\Omega_{i,j}$ and $Z$ are defined in (\ref{theorem2 eq1}).
	Then the observer and controller gains are given by: $K_i = M_{1i} Z_i^{-1}$, $L_{i1} = Q_i^{-1} M_{2i}$, $i = 1, 2, \dots, N$.
\end{theorem}
\begin{proof}
	On one hand, it has been proved in Theorem \ref{theorem2} that the composite closed-loop system (\ref{closed-sys2}) is stable with $H_\infty$ performance $\| G_{z_c d} \|_{\infty} < \gamma$, which gives (\ref{theorem3 eq1}). On the other hand, in (\ref{closed-sys2}) by setting $B K e = 0$, $F d = 0$ and $\Psi \Delta A x = 0$, the proof of the eigenvalue assignment part follows directly from Lemma \ref{lemma pole}, which yields (\ref{theorem3 eq2}) and (\ref{theorem3 eq3}).	\qed
\end{proof}

\section{Simulation results} \label{simulation}
\begin{figure}[t]
	\centering
	\includegraphics[scale=0.7]{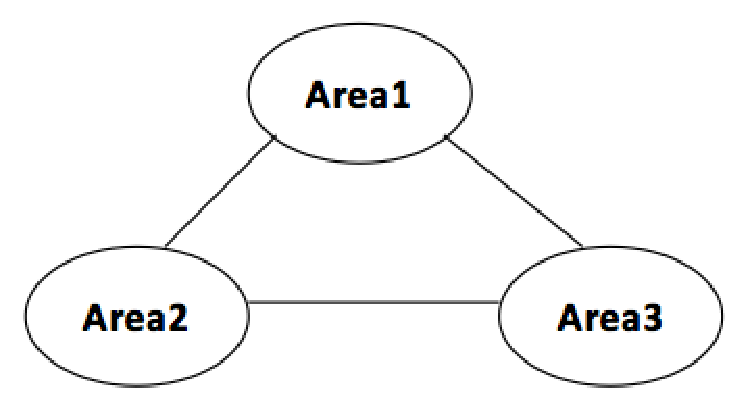}    
	\caption{Structure of the three-area power system}
	\label{fig8}
\end{figure}

A three-area interconnected power system (Figure \ref{fig8}) is used to demonstrate the effectiveness of the proposed integrated design strategy. To highlight its advantages, the integrated design is compared with the conventional decentralized separated designs \cite{mi2013decentralized,Su2017fault,rinaldi2017third,bernard2014decentralized,dong2012robust,zribi2005adaptive}, where all the local observers and controllers are designed separately and the area-interactions are treated as disturbances. For the sake of comparison, the two design strategies are solved by $H_\infty$ optimization using the same performance indices. The parameters of the power system are borrowed from \cite{jiang2012delay} and given in Table \ref{table 1}.
Two cases with and without eigenvalue assignment are simulated.   
\begin{table} [!htb]
	\renewcommand{\arraystretch}{1.1}
	\caption{Parameters of the three-area power system} \label{table 1}
	\centering
	{\tabcolsep10pt
		\begin{tabular}{@{}llllllll@{}}
			\hline
			& $T_{ch}$ & $T_{g}$ &  $R$ & $D$ & $M$ &  $\beta$ \\
			\hline
			\textbf{Area 1} & 0.3 & 0.1 &  0.05 & 1.0 & 10  &  1.0 \\
			\textbf{Area 2} & 0.4 & 0.17 &  0.05 & 1.5 & 12 &  1.0 \\
			\textbf{Area 3} & 0.35 & 0.2 &  0.05 & 1.8 & 12  &  1.0 \\
			\multicolumn{6}{c}{$T_{12}$=0.1986,~~~$T_{13}$=0.2148,~~~$T_{23}$=0.1830 } \\
			\hline
	\end{tabular}}
\end{table}

\subsection{Case 1: Without eigenvalue assignment}
For the proposed integrated and the separated design strategies, the parameters are chosen as $\gamma = 7.5$, $\varepsilon_1 = \varepsilon_2 = 1.0e-2$. Solving the LMI (\ref{theorem2 eq1}) gives the controller and observer gains of the integrated design strategy as follows:

$K_1 = [139.4504~    4.7855 ~   3.0247 ~  22.3814   ~ 6.4697] $,\\
$N_1 = 
\left[\AR{ccccc} 
-0.5018&         0&         0&   -0.0245&   -0.0094\\
-0.0181 &  -3.3333 &   3.3333 &  -0.0022 &  -0.0007\\
-0.0014  &       0 & -10.00  & -0.0002  & -0.0001\\
1.2881    &     0   &      0&  -39.890&   -0.5026\\
0.4969     &    0    &     0 &   0.4975 & -39.7847   
\AN\right] $,\\
$G_1 = \left[\AR{c}  0 \\ 0 \\ 10 \\ 0 \\ 0 \AN\right] ~H_1 = 
\left[\AR{ccc} 
1  &   0  &   0 \\
0   &  0   &  0 \\
0  &  0  &  0 \\
0  &   0  &   0 \\
0  &  0  &   0
\AN\right]$, ~
$L_1 = 
\left[\AR{ccc} 
0&    0.0245&    0.0094\\
0 &   0.0022 &   0.0007\\
-200.0000&    0.0002&    0.0001\\
2.5975 &  39.8900&    0.5026\\
1.0000  &  0.5025 &  39.7847 
\AN\right] $\\
%%%%%%%%%%%%%%%%%%%%%%
$K_2 = [250.7925 ~   8.6249~    5.7485 ~  39.5940 ~  10.9794] $,\\
$N_2 = 
\left[\AR{ccccc} 
-0.5015&         0&         0&   -0.0227&   -0.0094\\
-0.0286 &  -2.5000 &   2.5000 &  -0.0035 &  -0.0011\\
-0.0031  &       0  & -5.8824  & -0.0004  & -0.0001\\
1.1872    &     0    &     0&  -39.7763 &  -0.5026\\
0.4967     &    0     &    0 &   0.4974  &-39.7654   
\AN\right] $,\\
$G_2 = \left[\AR{c}  0 \\ 0 \\  5.8824 \\ 0 \\ 0 \AN\right], ~
H_2 = 
\left[\AR{ccc} 
1  &   0  &   0 \\
0   &  0   &  0 \\
0  &  0  &  0 \\
0  &   0  &   0 \\
0  &  0  &   0
\AN\right]$, ~
$L_2 = 
\left[\AR{ccc} 
0 &   0.0227 &   0.0094\\
0  &  0.0035  &  0.0011\\
-117.6471 &   0.0004 &   0.0001\\
2.3977 &  39.7763  &  0.5026\\
1.0000  &  0.5026   &39.7654
\AN\right]$,\\
%%%%%%%%%%%%%%%%%%555
$K_3 = [ 298.2493 ~   9.4408  ~  6.9871 ~  46.6206 ~  12.9404] $,\\
$N_3 = 
\left[\AR{ccccc} 
-0.5015   &      0 &        0 &  -0.0236&   -0.0094\\
-0.0245 &  -2.8571  &  2.8571  & -0.0031 &  -0.0010\\
-0.0042  &       0  & -5.0000  & -0.0005  & -0.0002\\
1.2382    &     0    &     0 & -39.7761 &  -0.5026\\
0.4968     &    0     &    0  &  0.4974 & -39.7653   
\AN\right] $,\\
$G_3 = \left[\AR{c}  0 \\ 0 \\  5 \\ 0 \\ 0 \AN\right],~
H_3 = 
\left[\AR{ccc} 
1  &   0  &   0 \\
0   &  0   &  0 \\
0  &  0  &  0 \\
0  &   0  &   0 \\
0  &  0  &   0
\AN\right] $,
$L_3 = 
\left[\AR{ccc} 
0  &  0.0236  &  0.0094\\
0   & 0.0031   & 0.0010\\
-100.0000 &   0.0005 &   0.0002\\
2.4995 &  39.7761 &   0.5026\\
1.0000 &   0.5026  & 39.7653
\AN\right].$\\

\begin{figure}[H]
	\centering
	\includegraphics[width=0.72\columnwidth]{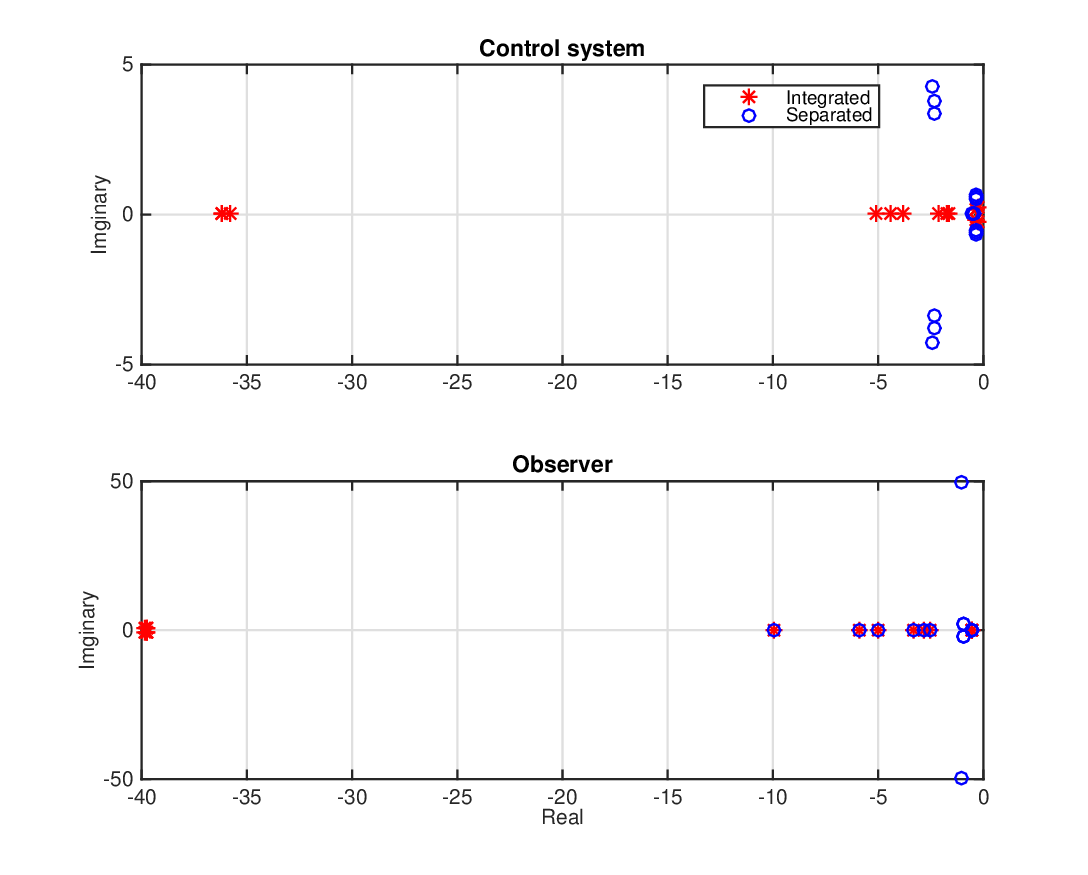}    
	\vspace{-2em}
	\caption{Eigenvalues of the control systems and observers using the integrated or separated strategies: Case 1}
	\label{fig4}
\end{figure}

\begin{figure}[!htb]
	\centering
	\includegraphics[width=0.70\columnwidth]{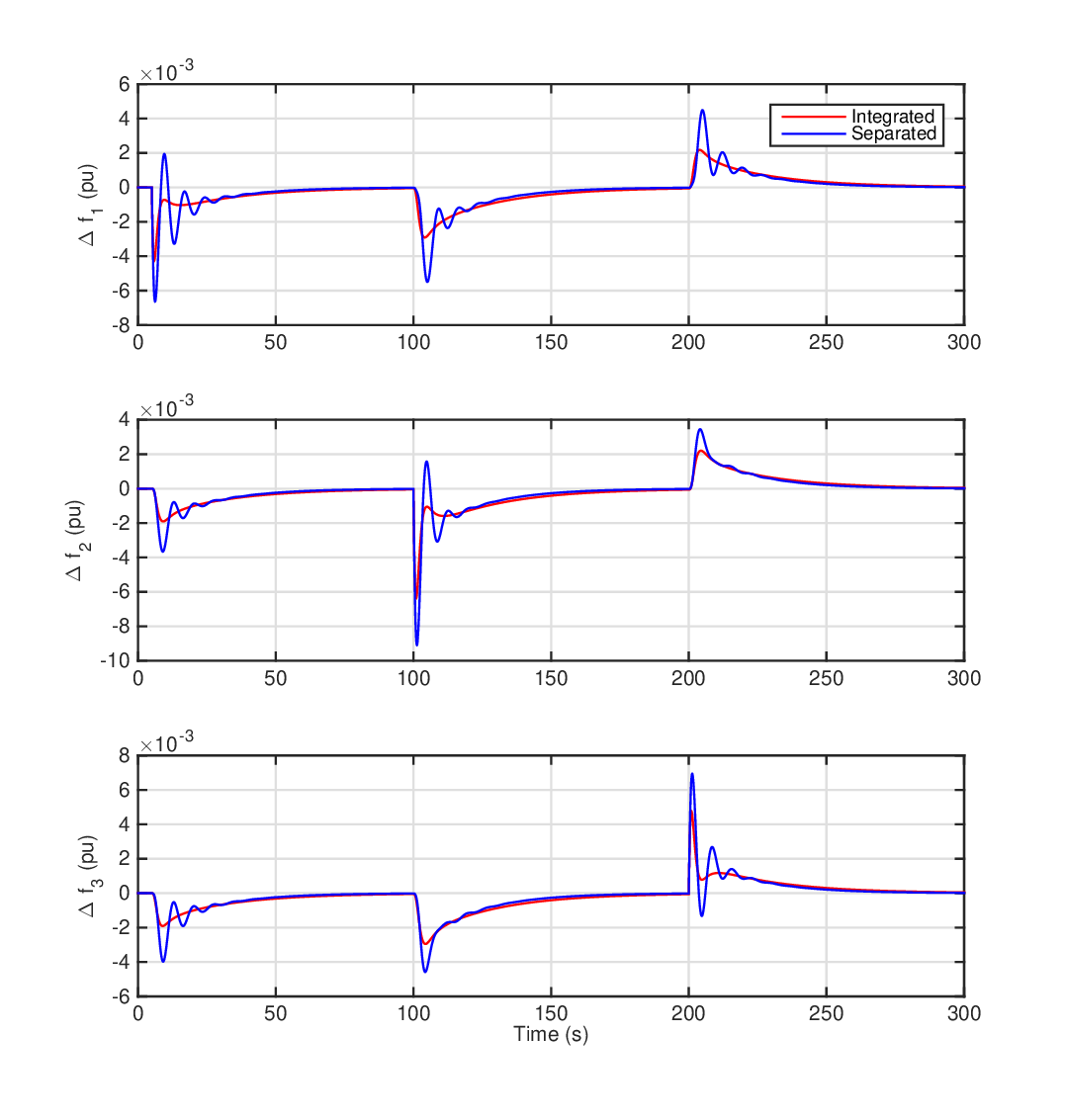}    
	\vspace{-2.5em}
	\caption{Frequency deviation: Case 1}
	\label{fig2}
\end{figure}

\begin{figure}[!htb]
	\centering
	\includegraphics[width=0.72\columnwidth]{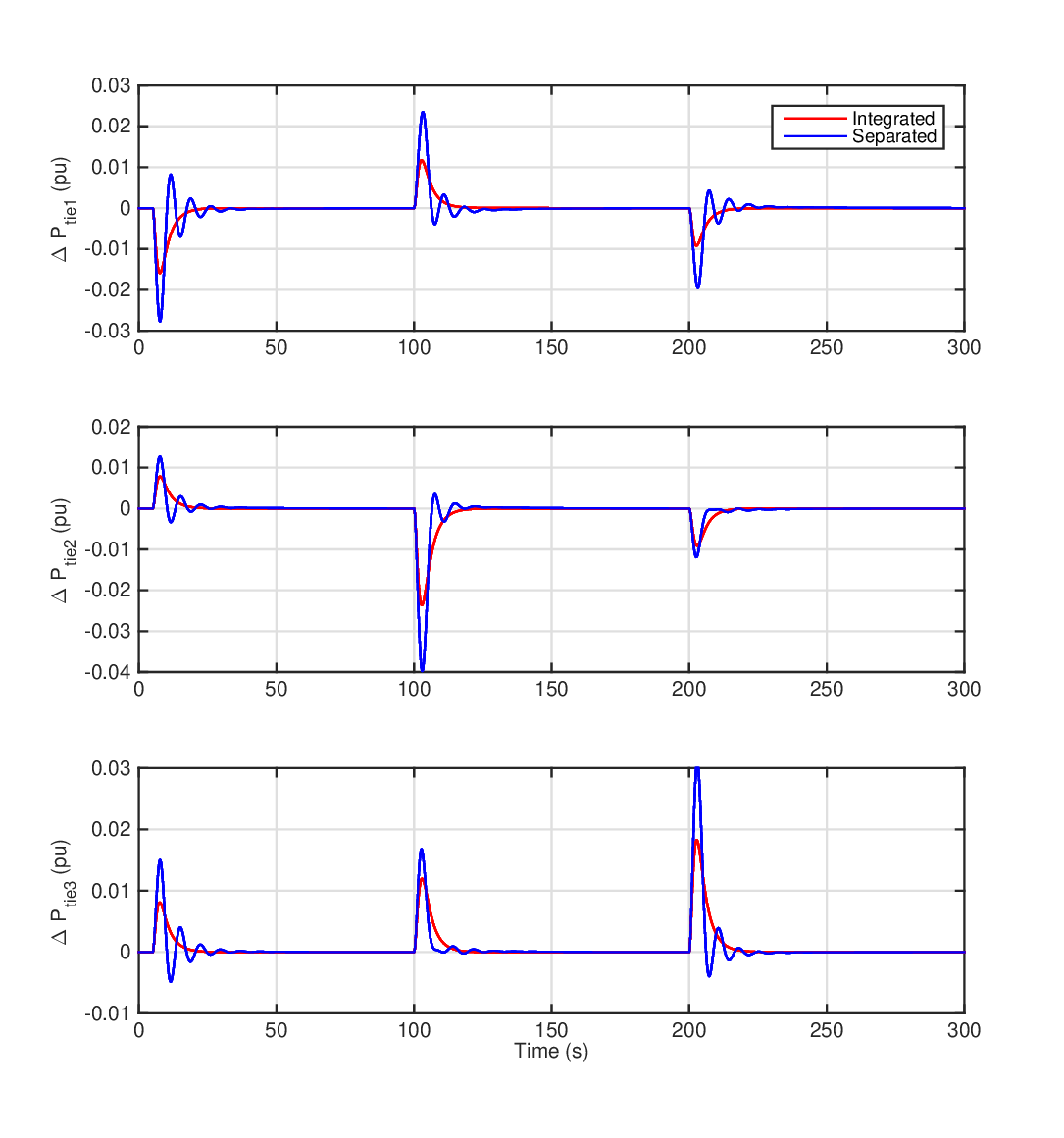}   
	\vspace{-2.5em} 
	\caption{Tie-line power flow deviation: Case 1}
	\label{fig3}
\end{figure}

The load disturbances are applied as +0.1 p.u. for area 1 at 5s, +0.15 p.u. for area 2 at 100s, and -0.12 p.u. for area 3 at 200s. It can be seen from Figure \ref{fig4} that compared with those of the separated design strategy, the eigenvalues of the control systems and observers of the integrated strategy are almost on the real axis. Hence, in Figures \ref{fig2} - \ref{fig3} the $\Delta f_i$ and $\Delta P_{tie_i}$, $i = 1, 2,3$, are damped to zero with less time, less and smaller oscillations by the integrated design strategy. 
%%%%%%%%%%%%%%%%%%%%%%%%%%%%%%%%%%%%%%%%%%%

\subsection{Case 2: With eigenvalue assignment}
For comparison, the proposed integrated and the separated design strategies use the same vertical strip regions for the eigenvalue assignment. As an example, in this simulation the strip regions for the control systems are chosen as: $a_{11} = a_{12} = a_{13} = -30$, $b_{11} = b_{12} = b_{13} = -1$, and for the observers are: $a_{21} = a_{22} = a_{23} = -40$, and $b_{21} = b_{22} = b_{23} = -2$. 
The other parameters are chosen as $\gamma = 1$, $\varepsilon_1 = 1.0e-4,\varepsilon_2 = 1.0e-3$, 
 Solving the LMIs (\ref{theorem3 eq1}), (\ref{theorem3 eq2}), and (\ref{theorem3 eq3}) together gives the controller and observer gains as follows:
 
 $K_1 = 10^3\times[2.3518 ~   0.0245~    0.0031~    1.6975~    1.9976] $,\\
 $N_1 = 
 \left[\AR{ccccc} 
 -3.7036&         0&         0&   -0.0055&   -0.0022\\
 28.8383 &  -3.3333 &   3.3333 &   0.0029 &   0.0022\\
 23.3571  &       0  &-10.00  &  0.0920  &  0.0369\\
 2.3141    &     0    &     0&  -34.3866&   -0.5073\\
 0.8880     &    0     &    0 &   0.4934 & -34.3672   
 \AN\right]$,\\
 $ G_1 = \left[\AR{c}  0 \\ 0 \\ 10 \\ 0 \\ 0 \AN\right], H_1 = 
 \left[\AR{ccc} 
 1  &   0  &   0 \\
 0   &  0   &  0 \\
 0  &  0  &  0 \\
 0  &   0  &   0 \\
 0  &  0  &   0
 \AN\right]$,
 $L_1 = 
 \left[\AR{ccc} 
 0&    0.0055 &   0.0022\\
 0 &  -0.0029  & -0.0022\\
 -200&   -0.0920 &  -0.0369\\
 2.5975&   34.3866&    0.5073\\
 1 &   0.5066 &  34.3672 
 \AN\right] $,\\
 %%%%%%%%%%%%%%%%%%%%%%
 $K_2 = 10^3\times[4.4557   ~ 0.0464~    0.0056~    3.1707~    3.6781] $,\\
 $N_2 = 
 \left[\AR{ccccc} 
 -3.3540&         0&         0&   -0.0061&   -0.0026\\
 19.7256 &  -2.500 &   2.5000 &   0.0066 &   0.0048\\
 25.2636  &       0  & -5.8824  &  0.0938  &  0.0407\\
 2.1695    &     0    &     0 & -34.3520 &  -0.5055\\
 0.9030     &    0     &    0  &  0.4941  &-34.3658   
 \AN\right] $,\\
 $G_2 = \left[\AR{c}  0 \\ 0 \\  5.8824 \\ 0 \\ 0 \AN\right],
 H_2 = 
 \left[\AR{ccc} 
 1  &   0  &   0 \\
 0   &  0   &  0 \\
 0  &  0  &  0 \\
 0  &   0  &   0 \\
 0  &  0  &   0
 \AN\right]$,
 $L_2 = 
 \left[\AR{ccc}
 0&    0.0061&    0.0026\\
 0 &  -0.0066 &  -0.0048\\
 -117.6471 &  -0.0938&   -0.0407\\
 2.3977&   34.3520&    0.5055\\
 1 &   0.5059 &  34.3658
 \AN\right]$,\\
 %%%%%%%%%%%%%%%%%%555
 $K_3 = 10^3\times[5.0295 ~   0.0493 ~   0.0068 ~   3.5844 ~   4.1438] $,\\
 $N_3 = 
 \left[\AR{ccccc} 
 -3.2931&         0&         0&   -0.0066&   -0.0027\\
 22.5433 &  -2.8571 &   2.8571 &   0.0077 &   0.0048\\
 23.7336  &       0  & -5.00  &  0.0942  &  0.0393\\
 2.2686    &     0    &     0 & -34.3516 &  -0.5059\\
 0.9053     &    0     &    0  &  0.4937  &-34.3654   
 \AN\right] $,\\
 $G_3 = \left[\AR{c}  0 \\ 0 \\  5 \\ 0 \\ 0 \AN\right], 
 H_3 = 
 \left[\AR{ccc} 
 1  &   0  &   0 \\
 0   &  0   &  0 \\
 0  &  0  &  0 \\
 0  &   0  &   0 \\
 0  &  0  &   0
 \AN\right]$,
 $L_3 = 
 \left[\AR{ccc} 
 0&    0.0066&    0.0027\\
 0 &  -0.0077 &  -0.0048\\
 -100.0000&   -0.0942&   -0.0393\\
 2.4995 &  34.3516&    0.5059\\
 1.0000  &  0.5063 &  34.3654
 \AN\right]. $\\
 
\begin{figure}[!htb]
	\centering
	\includegraphics[width=0.75\columnwidth]{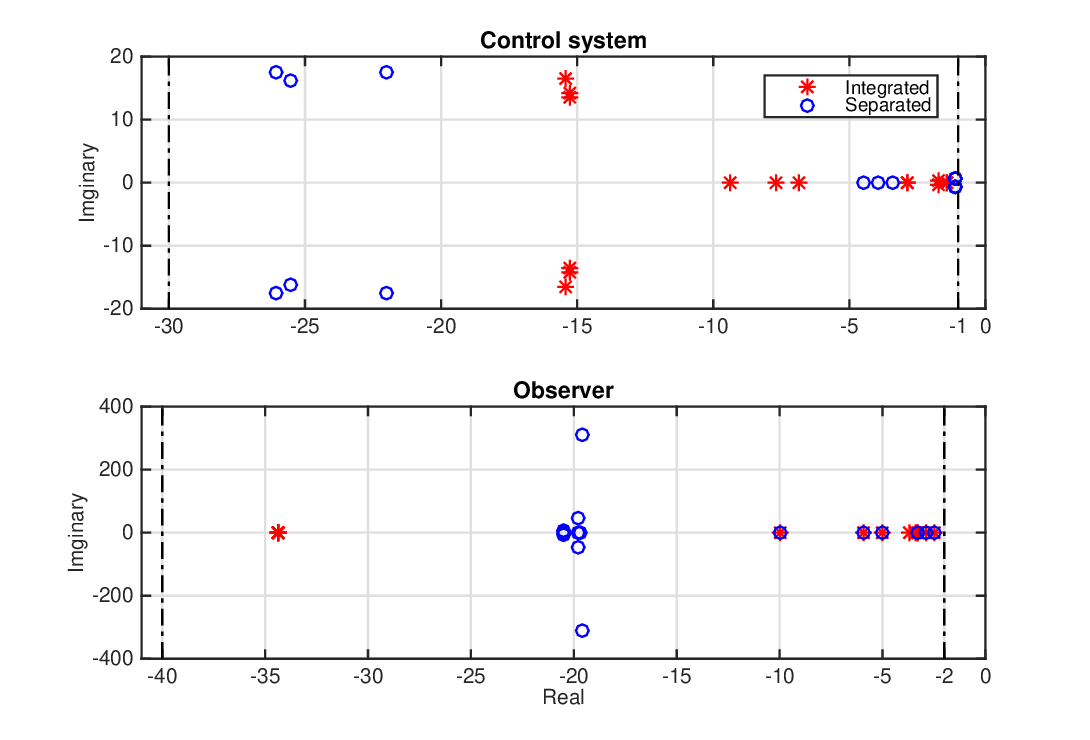}    
		\vspace{-1.5em}
	\caption{Eigenvalues of the control systems and observers using the integrated or separated strategies: Case 2}
	\label{fig7}
\end{figure}

\begin{figure}[!htb]
	\centering
	\includegraphics[width=0.75\columnwidth]{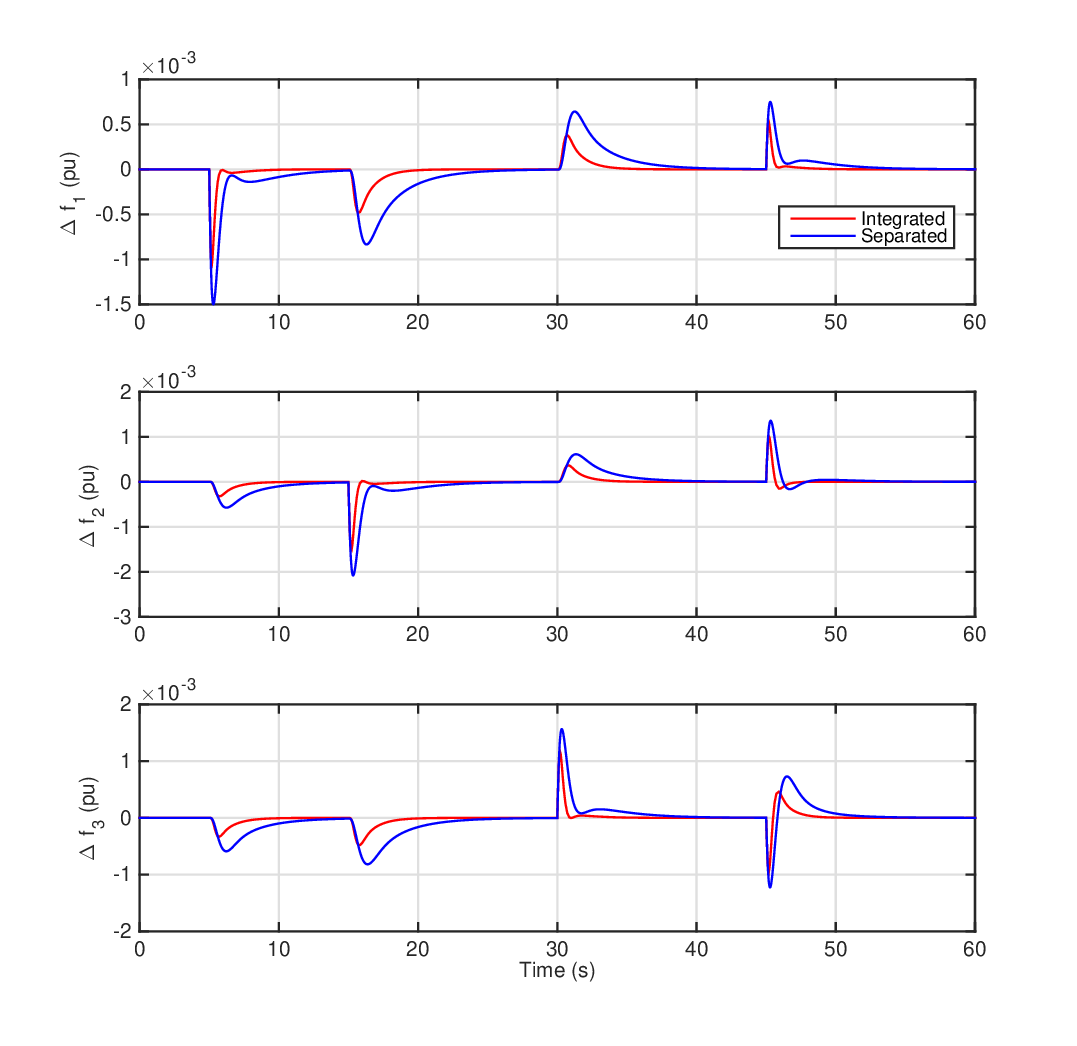}   
		\vspace{-2.5em} 
	\caption{Frequency deviation: Case 2}
	\label{fig5}
\end{figure}

\begin{figure}[!htb]
	\centering
	\includegraphics[width=0.74\columnwidth]{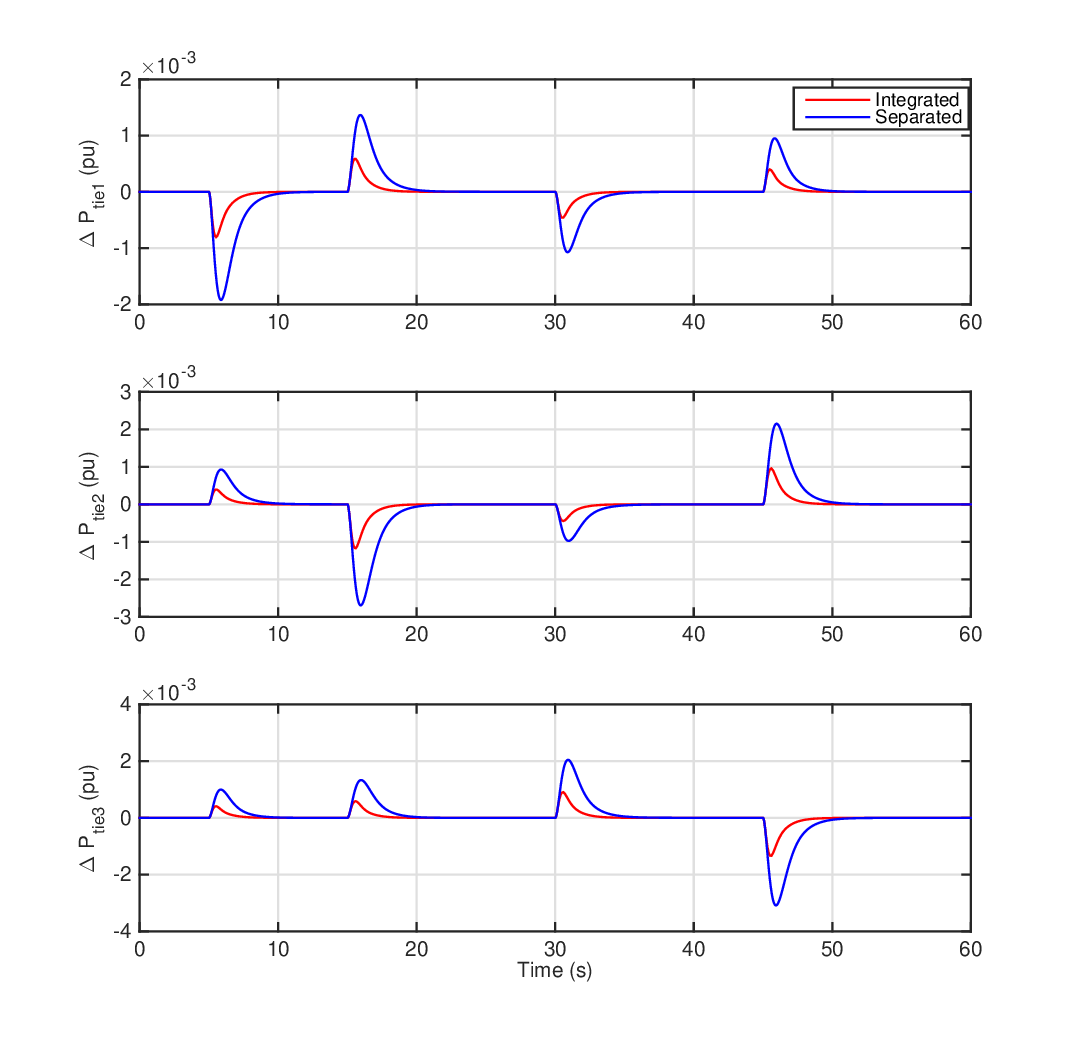}   
	\vspace{-2.5em} 
	\caption{Tie-line power flow deviation: Case 2}
	\label{fig6}
\end{figure}

The load disturbances are applied as +0.1 p.u. for area 1 at 5s, +0.15 p.u. for area 2 at 15s, -0.12 p.u. for area 3 at 30s, and +0.05 p.u., +0.05 p.u., -0.02 p.u for areas 1$\sim$3 at 45s. From Figure \ref{fig7} it can be seen that the eigenvalues of the control system and observers are all assigned to the prescribed regions. Although the eigenvalues of the integrated design strategy are not in better places than that of the separated one, it is shown in Figures \ref{fig5} - \ref{fig6} that the integrated strategy has better LFC performance than the separated one. Moreover, compared with Case 1, the results show that with eigenvalue assignment, the performances of both the strategies are greatly improved. 

\section{Conclusion} \label{conclusion}
This paper has proposed an integrated design strategy for designing the completely decentralized LFC robust observers and controllers, with considerations of the area-interactions between sub-areas and the bidirectional effects between the local observer and controller. In the proposed strategy, all the local observers and controllers are designed simultaneously off-line based on $H_\infty$ optimization using a one-step LMI method. Furthermore, the LMI regional eigenvalue assignment technique has been introduced and improved the transient performance of the closed-loop control system. Although the observers and controllers are decentralized in implementation, they have global optimized performance because of the integrated design strategy. A three-area power system has been simulated and validated the superiority of the proposed integrated design over the conventional decentralized designs under the same simulation conditions. 

%%%%%%%%%%%%%%%%%%%%%%%%%
\bibliographystyle{elsarticle-num}
\bibliography{References}

\begin{thebibliography}{10}
\expandafter\ifx\csname url\endcsname\relax
  \def\url#1{\texttt{#1}}\fi
\expandafter\ifx\csname urlprefix\endcsname\relax\def\urlprefix{URL }\fi
\expandafter\ifx\csname href\endcsname\relax
  \def\href#1#2{#2} \def\path#1{#1}\fi

\bibitem{kundur1994power}
P.~Kundur, N.~J. Balu, M.~G. Lauby, Power system stability and control, Vol.~7,
  New York, NY, USA: McGraw-Hill, 1994.

\bibitem{kumar2005recent}
P.~Kumar, D.~P. Kothari, Recent philosophies of automatic generation control
  strategies in power systems, IEEE Trans. Power Syst. 20~(1) (2005) 346--357.

\bibitem{shayeghi2009load}
H.~Shayeghi, H.~Shayanfar, A.~Jalili, Load frequency control strategies: A
  state-of-the-art survey for the researcher, Energy Convers. Manage. 50~(2)
  (2009) 344--353.

\bibitem{pandey2013literature}
S.~K. Pandey, S.~R. Mohanty, N.~Kishor, A literature survey on load frequency
  control for conventional and distribution generation power systems, Renew.
  Sustain. Energy Rev. 25 (2013) 318--334.

\bibitem{shankar2017comprehensive}
R.~Shankar, S.~Pradhan, K.~Chatterjee, R.~Mandal, A comprehensive state of the
  art literature survey on {LFC} mechanism for power system, Renew. Sustain.
  Energy Rev. 76 (2017) 1185--1207.

\bibitem{trinh2013quasi}
H.~Trinh, T.~Fernando, H.~H. Iu, K.~P. Wong, Quasi-decentralized functional
  observers for the {LFC} of interconnected power systems, IEEE Trans. Power
  Syst. 28~(3) (2013) 3513--3514.

\bibitem{pham2016load}
T.~N. Pham, H.~Trinh, Load frequency control of power systems with electric
  vehicles and diverse transmission links using distributed functional
  observers, IEEE Trans. Smart Grid 7~(1) (2016) 238--252.

\bibitem{bevrani2004robust}
H.~Bevrani, Y.~Mitani, K.~Tsuji, Robust decentralised load-frequency control
  using an iterative linear matrix inequalities algorithm, IEE Proc., Gener.
  Transm. Distrib. 151~(3) (2004) 347--354.

\bibitem{khodabakhshian2008new}
A.~Khodabakhshian, M.~Edrisi, A new robust {PID} load frequency controller,
  Control Eng. Pract. 16~(9) (2008) 1069--1080.

\bibitem{tan2010unified}
W.~Tan, Unified tuning of {PID} load frequency controller for power systems via
  {IMC}, IEEE Trans. Power Syst. 25~(1) (2010) 341--350.

\bibitem{rerkpreedapong2003robust}
D.~Rerkpreedapong, A.~Hasanovic, A.~Feliachi, Robust load frequency control
  using genetic algorithms and linear matrix inequalities, IEEE Trans. Power
  Syst. 18~(2) (2003) 855--861.

\bibitem{juang2006load}
C.-F. Juang, C.-F. Lu, Load-frequency control by hybrid evolutionary fuzzy {PI}
  controller, IEE Proc., Gener. Transm. Distrib. 153~(2) (2006) 196--204.

\bibitem{tan2012robust}
W.~Tan, H.~Zhou, Robust analysis of decentralized load frequency control for
  multi-area power systems, Int. J. Elect. Power Energy Syst. 43~(1) (2012)
  996--1005.

\bibitem{mi2013decentralized}
Y.~Mi, Y.~Fu, C.~Wang, P.~Wang, Decentralized sliding mode load frequency
  control for multi-area power systems, IEEE Trans. Power Syst. 28~(4) (2013)
  4301--4309.

\bibitem{Su2017fault}
S.~Xiaojie, L.~Xinxin, S.~Yongduan, Fault-tolerant control of multi-area power
  systems via sliding mode observer technique, IEEE/ASME Trans. on Mechatronics
  (2017).
\newblock \href {https://doi.org/10.1109/TMECH.2017.2718109}
  {\path{doi:10.1109/TMECH.2017.2718109}}.

\bibitem{rinaldi2017third}
G.~Rinaldi, M.~Cucuzzella, A.~Ferrara, Third-order sliding mode observer-based
  approach for distributed optimal load frequency control, IEEE Control Systems
  Letters 1~(2) (2017) 215--220.

\bibitem{bernard2014decentralized}
M.~Z. Bernard, T.~H. Mohamed, Y.~S. Qudaih, Y.~Mitani, Decentralized load
  frequency control in an interconnected power system using coefficient diagram
  method, Int. J. Elect. Power Energy Syst. 63 (2014) 165--172.

\bibitem{dong2012robust}
L.~Dong, Y.~Zhang, Z.~Gao, A robust decentralized load frequency controller for
  interconnected power systems, ISA Trans. 51~(3) (2012) 410--419.

\bibitem{zribi2005adaptive}
M.~Zribi, M.~Al-Rashed, M.~Alrifai, Adaptive decentralized load frequency
  control of multi-area power systems, Int. J. Elect. Power Energy Syst. 27~(8)
  (2005) 575--583.

\bibitem{alrifai2011decentralized}
M.~T. Alrifai, M.~F. Hassan, M.~Zribi, Decentralized load frequency controller
  for a multi-area interconnected power system, Int. J. Elect. Power Energy
  Syst. 33~(2) (2011) 198--209.

\bibitem{kazemi2002decentralized}
M.~H. Kazemi, M.~Karrari, M.~B. Menhaj, Decentralized robust adaptive-output
  feedback controller for power system load frequency control, Electr. Eng.
  84~(2) (2002) 75--83.

\bibitem{rerkpreedapong2002decentralized}
D.~Rerkpreedapong, A.~Feliachi, Decentralized load frequency control for load
  following services, in: IEEE Power Engineering Society Winter Meeting,
  Vol.~2, IEEE, 2002, pp. 1252--1257.

\bibitem{feliachi1987optimal}
A.~Feliachi, Optimal decentralized load frequency control, IEEE Trans. Power
  Syst. 2~(2) (1987) 379--385.

\bibitem{jiang2012delay}
L.~Jiang, W.~Yao, Q.~Wu, J.~Wen, S.~Cheng, Delay-dependent stability for load
  frequency control with constant and time-varying delays, IEEE Trans. Power
  Syst. 27~(2) (2012) 932--941.

\bibitem{patton1997observer}
R.~Patton, J.~Chen, Observer-based fault detection and isolation: Robustness
  and applications, Control Eng. Pract. 5~(5) (1997) 671--682.

\bibitem{boyd1994linear}
S.~Boyd, L.~El~Ghaoui, E.~Feron, V.~Balakrishnan, Linear matrix inequalities in
  system and control theory, SIAM, 1994.

\bibitem{chilali1996h}
M.~Chilali, P.~Gahinet, {$H_\infty$} design with pole placement constraints: an
  {LMI} approach, IEEE Trans. on Autom. Control 41~(3) (1996) 358--367.

\end{thebibliography}

\end{document}